\documentclass[apj]{emulateapj}
\usepackage{times}
\usepackage{lscape}
\bibpunct[; ]{(}{)}{;}{a}{}{,}

\renewcommand{\ion}[2]{#1$\;${\scshape{#2}}}

\newcommand{\feii}{\mbox{\ion{Fe}{ii}}}

\newcommand{\oiii}{\mbox{[\ion{O}{iii}]}}
\newcommand{\ha}{\mbox{H$\alpha$}}

\shortauthors{Desroches, Greene \& Ho}
\shorttitle{X-ray Properties of IMBHs}

\begin{document}


\slugcomment{revised \today; submitted to the {\it Astrophysical Journal}}

\title{X-ray Properties of Intermediate-Mass Black Holes in Active Galaxies. II. X-ray-Bright Accretion and Possible Evidence for Slim Disks}
\author{Louis-Benoit Desroches\altaffilmark{1}, Jenny E. Greene\altaffilmark{2,3}, and Luis C. Ho\altaffilmark{4}}

\altaffiltext{1}{Department of Astronomy, University of California, Berkeley,
CA 94720-3411; louis@astro.berkeley.edu}
\altaffiltext{2}{Department of Astrophysical Sciences, Princeton University, Princeton, NJ 08544; jgreene@astro.princeton.edu}
\altaffiltext{3}{Hubble, Princeton-Carnegie Fellow}
\altaffiltext{4}{The Observatories of the Carnegie Institution of Washington, 
813 Santa Barbara St., Pasadena, CA 91101; lho@ociw.edu}


\begin{abstract}
We present X-ray properties of optically-selected intermediate-mass ($\sim10^5$--$10^6 M_\odot$) black holes (BHs) in active galaxies (AGNs), using data from the {\em Chandra} X-Ray Observatory. Our observations are a continuation of a pilot study by \cite{GH07a}. Of the 8 objects observed, 5 are detected with X-ray luminosities in the range $L_{\rm 0.5-2 \ keV} = 10^{41}$--$10^{43}$ erg s$^{-1}$, consistent with the previously observed sample. Objects with enough counts to extract a spectrum are well fit by an absorbed power law. We continue to find a range of soft photon indices $1 < \Gamma_s < 2.7$, where $N(E) \propto E^{-\Gamma_s}$, consistent with previous AGN studies, but generally flatter than other narrow-line Seyfert 1 active nuclei (NLS1s). 
The soft photon index correlates strongly with X-ray luminosity and Eddington ratio, but does not depend on BH mass. There is no justification for the inclusion of any additional components, such as a soft excess, although this may be a function of the relative inefficiency of detecting counts above 2 keV in these relatively shallow observations. As a whole, the X-ray-to-optical spectral slope $\alpha_{\rm ox}$ is flatter than in more massive systems, even other NLS1s. Only X-ray-selected NLS1s with very high Eddington ratios share a similar $\alpha_{\rm ox}$. This is suggestive of a physical change in the accretion structure at low masses and at very high accretion rates, possibly due to the onset of slim disks. Although the detailed physical explanation for the X-ray loudness of these intermediate-mass BHs is not certain, it is very striking that targets selected on the basis of optical properties should be so distinctly offset in their broader spectral energy distributions.
\end{abstract}



\section{INTRODUCTION}

Supermassive black holes (BHs), with masses of $\sim 10^6$--$10^9 \ M_\odot$, exist at the center of nearly all elliptical galaxies and galaxy bulges, as determined from stellar and gas dynamics, and from the presence of actively accreting galactic centers (active galactic nuclei; AGNs). An important problem  yet to be resolved in cosmological galaxy evolution is understanding the creation and growth of ``seed'' BHs. Stellar-mass BHs, the end product of massive stars, have masses of only $\approx 10 M_\odot$, leaving a gap of 5 orders of magnitude in BH mass. BHs in this unknown region are often dubbed intermediate-mass BHs (or low-mass galactic BHs). The recent discoveries of previously unknown $\approx$10$^5 \ M_\odot$ galactic BHs are beginning to constrain formation and evolution models of such seed BHs. NGC\,4395, a bulgeless late-type spiral galaxy, and POX\,52, a spheroidal galaxy, both contain Seyfert 1 AGNs with BH masses estimated to be $\approx10^5 \ M_\odot$ \citep{FH03,Peterson05,Barth04}.

BH mass ($M_{\rm BH}$) correlates strongly with various properties of spheroidal systems, such as luminosity \citep{KR95} and stellar velocity dispersion \citep{Gebhardt00,FM00,Tremaine02}. The evolution of bulges and the growth of BHs are thus likely coupled, although the possible existence of nuclear BHs in bulgeless or nearly bulgeless spiral galaxies \citep{DH09} suggests that bulges are not a necessary condition for BH growth. Nuclear intermediate-mass BHs are unfortunately very difficult to detect; the gravitational sphere of influence of a $10^5 M_\odot$ BH is unresolvable beyond the Local Group, even with the {\em Hubble Space Telescope}. We can rely on AGN signatures, however, to signify the presence of a BH, and use the observed broad emission lines to estimate BH mass \citep{GH05}.  \cite{GH04} systemically searched the First Data Release of the Sloan Digital Sky Survey (SDSS) and found 19 galaxies with BH estimates of $\lesssim 10^6 \ M_\odot$, which forms the parent sample of this study.

The homogenous selection of the \citeauthor{GH04} sample allows for important broadband, multiwavelength investigations, to determine how spectral properties change with BH mass in this intermediate-mass regime. These objects were found to be radio-faint using the Very Large Array \citep{GHU06}. The X-ray luminosity, from a pilot study of this sample (\citealt{GH07a}; hereafter referred to as Paper I), ranges from $L_{\rm 0.5-2 \ keV} \approx 10^{41}$ to $10^{43}$ erg s$^{-1}$. Here we present the rest of the X-ray results for the remaining objects in the sample. 

The observations and data analysis are discussed in \S~\ref{sec:sample}. We present our results and discuss physical connections to other AGNs in \S~\ref{sec:results}. Finally, we summarize our findings in \S~\ref{sec:sum}. We assume a cosmology such that $H_0 = 71$ km s$^{-1}$ Mpc$^{-1}$, $ \Omega_m = 0.27$, and $\Omega_\Lambda = 0.75$ \citep{Spergel03}.

\section{X-RAY SAMPLE AND DATA ANALYSIS}
\label{sec:sample}

We observed 8 intermediate-mass BHs from \cite{GH04}, which were not observed in Paper I, with the Advanced CCD Imaging Spectrometer (ACIS; \citealt{Garmire03}) on board {\em Chandra} \citep{Weisskopf96}.  The observations were obtained during Guest Observer Cycle 8 between 2007 March and 2007 August. As in Paper I, images were obtained at the aim point of the S3 CCD in faint mode. We once again read out only 1/8 of the chip, with a minimum read-out time of 0.4 s, to reduce the effects of pile-up. Effective exposure times ranged from  4.98 ks to 5.49 ks.

We use standard type 2 event files, processed by the {\em Chandra} X-Ray Center, for further analysis. We use the CIAO ({\em Chandra} Interactive Analysis of Observations) task {\tt celldetect}, with default parameters, to automatically detect and extract centroid positions for each source. Of the 8 objects observed, 5 are detected. One source (GH03) required setting the minimum signal-to-noise (S/N) threshold down to 2 from the default of 3 in order to detect it. The on-axis point-spread function (PSF) of {\em Chandra} contains 95\% of the encircled energy within 1\arcsec, and so we adopt a 2\arcsec \ radius aperture and extract background-subtracted counts using the CIAO task {\tt dmextract}. The background regions are annuli of inner radius 7\arcsec \ and outer radius 15\arcsec. Counts are extracted in two bands: the soft band (0.5--2 keV; $C_s$) and the hard band (2--8 keV; $C_h$). As in Paper I, 1\arcsec \ extractions are consistent with encircling 95\% of the energy, justifying our choice of a 2\arcsec \ aperture. For those objects without a detection, we measured the background counts in the 2\arcsec \ aperture, and determined\footnote{See the online \texttt{celldetect} documentation for a detailed description of how the S/N ratio is calculated.} the counts necessary for a theoretical source to be detected by {\tt celldetect} with our required minimum S/N. 

We calculate a ``hardness ratio'' for our detected objects, defined as $H \equiv (C_h-C_s)/(C_h+C_s)$, as a rough estimate of the spectral shape. We can then use this ratio to infer a soft photon index $\Gamma_s$, where $N(E) \propto E^{-\Gamma_s}$, as described by \cite{Gallagher05} and used in Paper I. To accomplish this, we build artificial spectra with known photon indices and Galactic absorption, and then ``observe'' them with the same instrumental response as the observations. The response is characterized by the redistribution matrix file (RMF), which modifies the input energy spectrum into the observed distribution of pulse heights due to finite energy resolution, and the auxiliary response file (ARF), which modifies the input spectrum due to the effective area and quantum efficiency of the detectors. The task {\tt fakeit} within the spectral-fitting package {\tt XSPEC} \citep{Arnaud96} is used to generate artificial spectra with photon indices ranging from $\Gamma=1$ to $3$ and with the same Galactic absorption as the true observation, determined from \cite{DL90} using WebPIMMS\footnote{\tt http://heasarc.gsfc.nasa.gov/Tools/w3pimms.html}. The hardness ratio is then measured from these artificial spectra and compared to the true ratio. By matching the hardness ratios, we can thus infer a photon index $\Gamma_{\rm HR}$. This photon index and known neutral column density $N_{\rm H}$ are then used with WebPIMMS to calculate fluxes from the observed count rates. Results are presented in Table~\ref{tab:prop}.
We note that because we do not include an intrinsic neutral column in our calculations, our photon indices represent lower limits to the true value. Given our acceptable spectral fits, however, we see no compelling reason to suspect a significant contribution from such a component (see below and Paper I). 

\subsection{Spectral Fitting}

Two of our observed targets (GH12 and GH17) have enough counts ($>$200) for a reliable spectral fit. We use the same aperture and background region as above to extract a spectrum using the CIAO task {\tt psextract}, which also generates the appropriate RMF and ARF files. We select a minimum of 20 counts for each energy bin to permit the use of $\chi^2$ statistics. We limit our analysis to 0.3--5 keV and 0.5-5 keV for GH12 and GH17, respectively, to avoid large uncertainties due to detector response and low counts.

\begin{figure}
\epsscale{1.0}
\plottwo{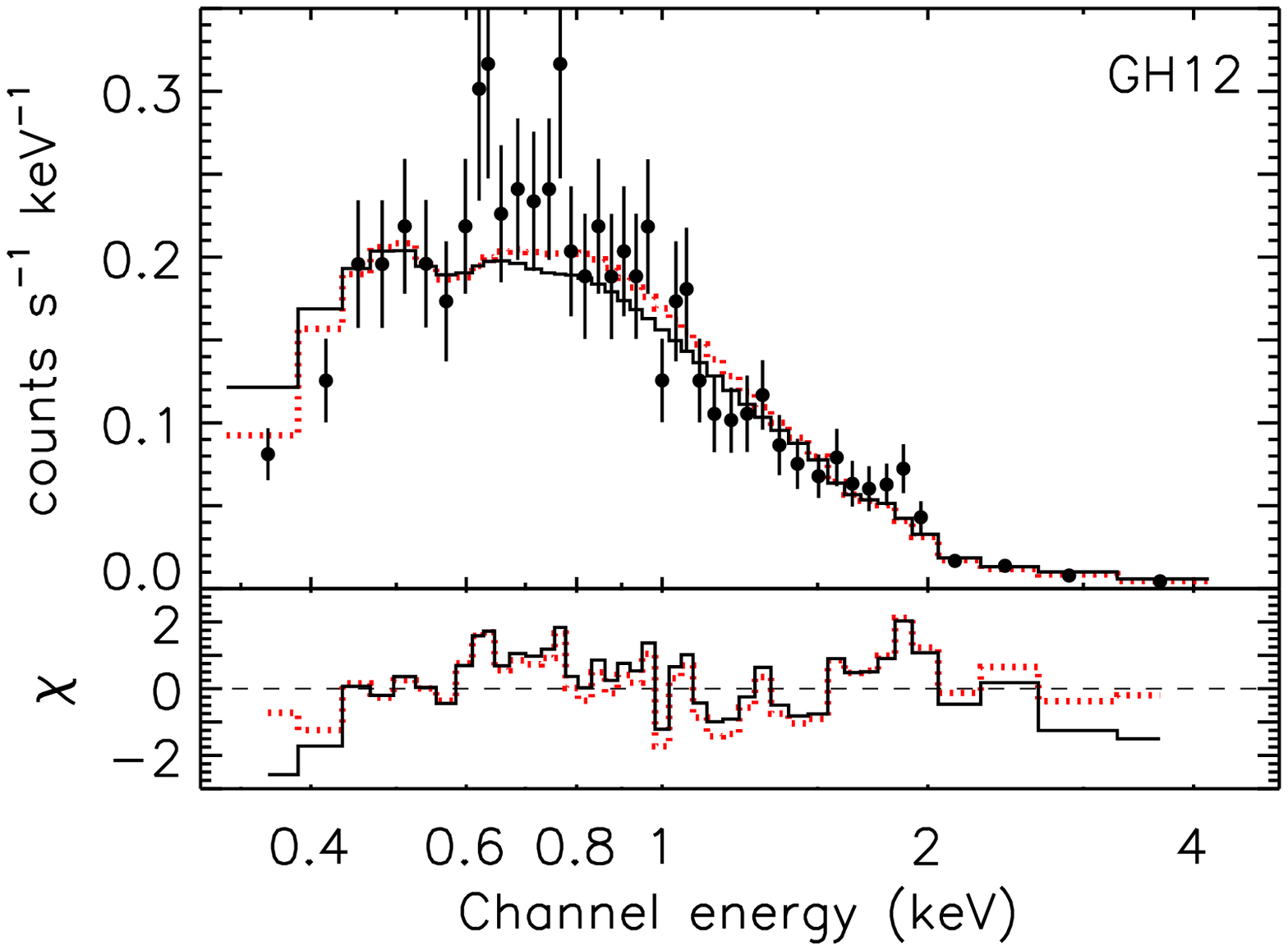}{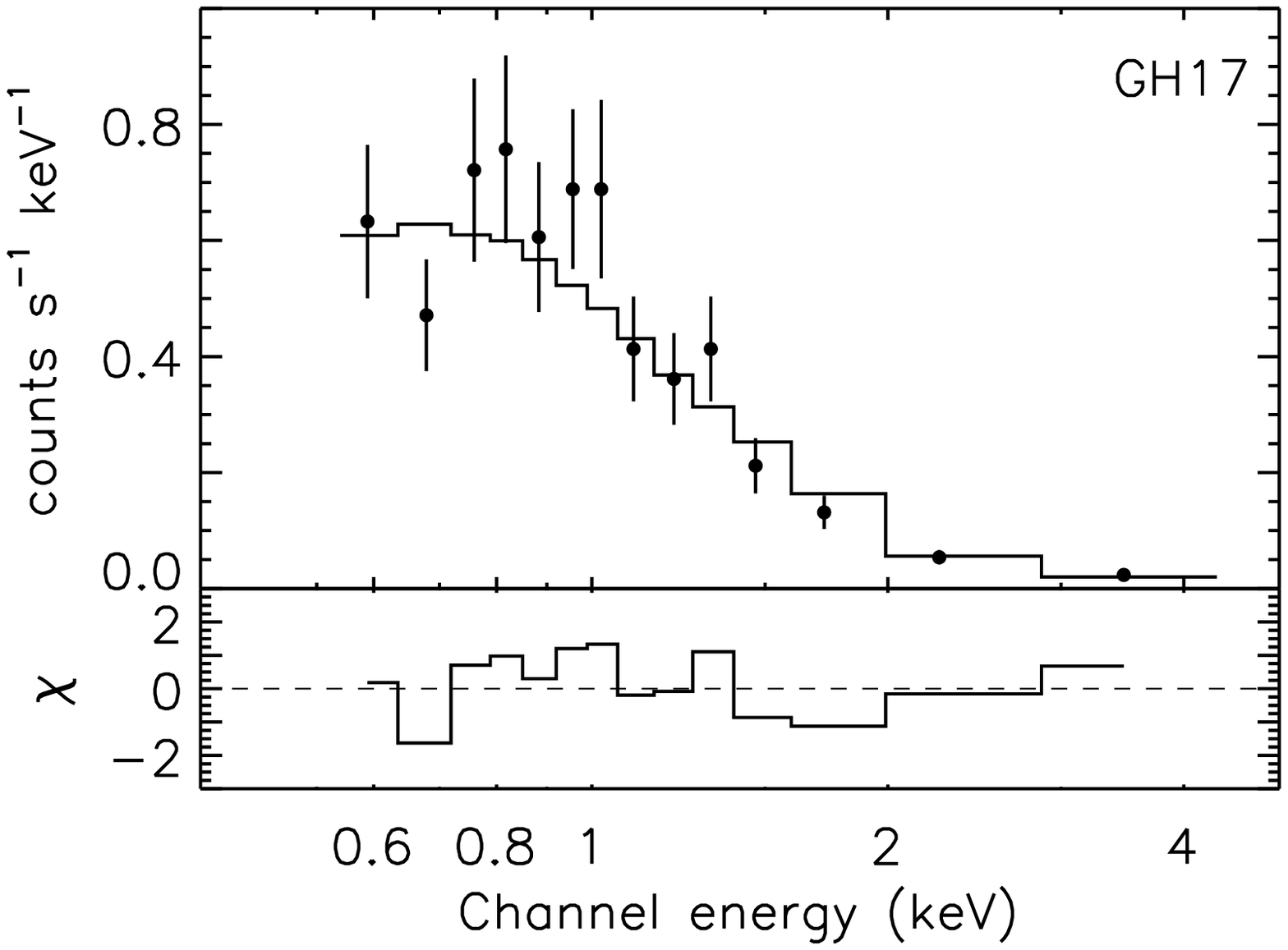}
\caption{Extracted X-ray spectra for objects with $> 200$ counts. Energy bins are chosen to have a minimum of 20 counts. Also plotted are absorbed power-law models ({\it solid black line}) with neutral column density $N_{\rm H}$ fixed to the value in Table~\ref{tab:prop} \citep{DL90}. For GH12 we also plot an absorbed power-law model with $N_{\rm H}$ as a free parameter ({\it dotted red line}), which results in a marginally better fit. Fit parameters are given in Table~\ref{tab:spectra}. The bottom panel in each plot shows the residuals for each model normalized by the $1\sigma$ uncertainty in the measurement.}
\label{fig:spec}
\end{figure}

Using {\tt XSPEC}, we fit a simple absorbed power-law model to each spectrum. The value for the absorption is fixed to the Galactic value \citep{DL90}. These fits are shown in Figure~\ref{fig:spec} and Table~\ref{tab:spectra}, with 90\% confidence errors quoted. In both cases, the reduced $\chi_\nu^2$ is consistent with 1, indicating a reliable fit with no additional components necessary. The photon indices $\Gamma_{\rm HR}$ and $\Gamma_s$ are similar in both cases. We also performed fits with the absorption parameter allowed to vary. For GH12, the resulting fit differs by more than the 90\% confidence errors. The new fit parameters are $N_{\rm H} = (8 \pm 2) \times 10^{20}$ cm$^{-2}$, $\Gamma_s = 2.77 \pm 0.15$, normalization $= (3.8 \pm 0.7) \times 10^{-4}$ photons s$^{-1}$ keV$^{-1}$ at 1 keV, and $\chi_\nu^2=0.88$.  The differing neutral column density between the two fits may be tentative evidence for an intrinsic absorber, although with $\chi_\nu^2$ formally less than 1, we may be simply overfitting the data. In the case of GH17, allowing $N_{\rm H}$ to vary results in a similar fit (within uncertainties). In neither case do the data have sufficient depth nor spectral coverage to model any additional components, if present, such as a soft thermal excess. Our model is, as a result, oversimplified but acceptable given the spectral information available.

In Paper I, only GH04 displayed marginal evidence for a soft excess. 
The absence of evidence for soft excesses in our data may be a function of the relatively soft energies detectable by {\em Chandra} in short exposures, in addition to our limited S/N. 
\cite{Miniutti09} observed 4 objects in our sample, detected by both {\em Chandra} and {\em ROSAT}, with {\em XMM-Newton}. They found an apparent break in the X-ray spectra of all 4 objects at $\approx 2$ keV. An absorbed power-law fit, using only the hard 2--10 keV counts, produces a very noticeable soft excess when extrapolated to softer energies. The hard photon index $\Gamma_h$ (2--10 keV) is also much flatter than our observed soft photon index, with $\langle \Gamma_h \rangle = 1.76$. Thus, with long exposures capable of detecting significant hard X-ray counts, these intermediate-mass BHs appear to behave very similarly to other radio-quiet, type 1 Seyferts and quasars which exhibit very pronounced soft excesses below 1 keV \citep{BBF96}. The physical explanation for such an excess is still unclear \citep{Miniutti09}.

\section{RESULTS AND DISCUSSION}
\label{sec:results}

\subsection{Narrow-Line Seyfert 1 Galaxies}
\label{sec:nls1}

The \cite{GH04} sample can be classified as narrow-line Seyfert 1 (NLS1) galaxies, a subclass of AGNs, based on the width of the broad permitted lines, in particular FWHM$_{\rm H \beta} < 2000$ km s$^{-1}$. NLS1s are thought to be intermediate-mass BHs radiating at high Eddington ratios \citep{PDO95}, the ``narrow'' broad lines a result of the small virial velocities associated with the intermediate-mass BH. Indeed the $M_{\rm BH}$ and Eddington ratio ($L_{\rm bol}/L_{\rm Edd}$) estimates from \cite{GH04} support the picture that this sample is at the low-mass end of the classical NLS1 subclass. \cite{Collin06} postulate, however, that NLS1s arise instead from high-mass BH systems observed at high inclination, reproducing the ``narrow'' lines. As noted in Paper I, we cannot rule this out, but the agreement of the \citeauthor{GH04} objects with the extrapolation of the $M_{\rm BH}$--$\sigma_\star$ relation suggests that these are true intermediate-mass BHs \citep{BGH05,GH06}. The host galaxies are similarly low-mass and low-luminosity \citep{GHB08}. Although some optical spectroscopic properties, such as the strength of \feii \ and \oiii \ lines, differ between classical NLS1s and the \citeauthor{GH04} sample, both groups tend to be exceptionally radio-quiet \citep{GHU06}.


Characteristic NLS1 properties, such as FWHM$_{\rm H\beta}$ and the soft X-ray photon index, have been suggested to depend on the BH mass. \cite{BBF96} and \cite{Porquet04} have argued in particular that low BH masses are necessary for steep photon indices. In Paper I, however, the observed sample covered a range of $600 \leq$ FWHM$_{\rm H\beta}$ $\leq 1800$ km s$^{-1}$ and $1 \leq \Gamma_s \leq 3$, with a mean $
\Gamma_s = 2.1\pm0.2$,
flatter than the {\em XMM-Newton} PG quasar sample of \cite{Porquet04} despite having {\em lower} BH masses ($10^5 < M_{\rm BH}/M_\odot < 10^{6.5}$). Including our current sample,  the mean becomes  $\Gamma_s= 2.2 \pm 0.1$
compared to $\Gamma_s = 2.6 \pm 0.1$ 
from \cite{Porquet04} and $\Gamma = 2.58 \pm 0.05$ measured for general {\em R\"{o}ntgensatellit} ({\em ROSAT}) samples \citep{Yuan98}. As was determined in Paper I, this clearly signifies that low BH mass is not sufficient for a steep soft X-ray power law. Unlike in classic NLS1 samples, the \cite{GH04} sample is not soft-X-ray selected, but rather selected to have low BH mass, thus leading to a wider distribution of Eddington ratios.
The high $\Gamma_s$, low-BH-mass objects from \cite{Porquet04}, however, are also the objects with the highest Eddington ratios.

\begin{figure}
\epsscale{1.0}
\plotone{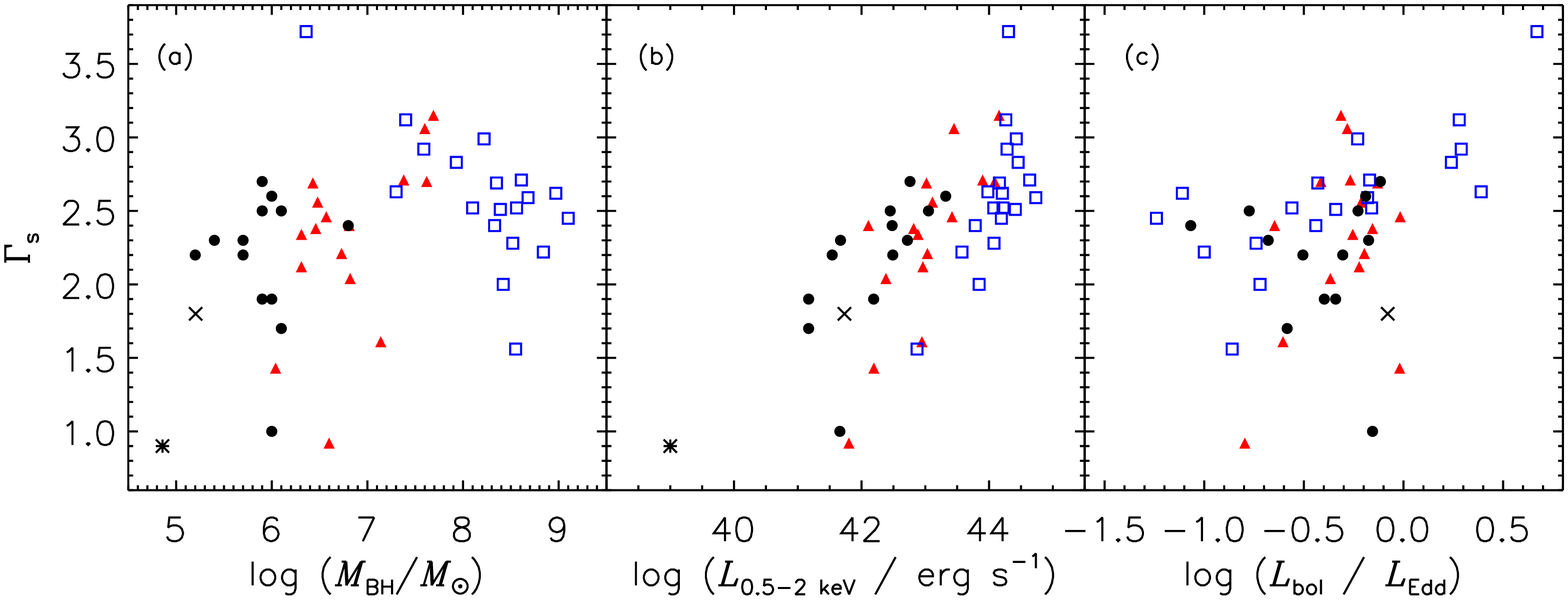}
\caption{(a) $\Gamma_s$ (0.5--2 keV) vs. $M_{\rm BH}$. The \cite{GH04} objects detected by {\em Chandra} ({\it black circles}) populate the intermediate-mass regime, but show that $\Gamma_s$ is not strongly correlated with BH mass. We also include optically selected, X-ray-weak NLS1s \cite[{\it red triangles}]{WMP04}, PG quasars observed with {\em XMM-Newton} \cite[{\it blue open squares}]{Porquet04}, NGC\,4395 \citep[{\it asterisk}]{Moran05}, and POX\,52 \citep[{\it cross}]{Thornton08}. (b) $\Gamma_s$ vs. $L_{\rm 0.5-2 \ keV}$. This relation exhibits the strongest correlation. (c) $\Gamma_s$ vs. $L_{\rm bol} / L_{\rm Edd}$. $L_{\rm bol}$ is estimated from optical observations only, to avoid any potential secondary correlations between X-ray luminosity and slope. We use $L_{\rm H\alpha}$ to estimate $L_{5100\AA}$ \citep{GH05,GH07b} for our {\em Chandra} sample, and then assume $L_{\rm bol} = 9 L_{5100\AA}$ \citep{Ho08}.
}
\label{fig:gammaplots}
\end{figure}

With X-ray properties of the full \cite{GH04} sample, we can revisit the possible correlations of $\Gamma_s$ with other parameters, such as $L_{\rm 0.5-2 \ keV}$, and $L_{\rm bol}/L_{\rm Edd}$. This is shown in Figure~\ref{fig:gammaplots}. As we have already noted, $M_{\rm BH}$ by itself is a poor indicator of soft photon index. The X-ray luminosity $L_{\rm 0.5-2 \ keV}$ and Eddington ratio continue to be significantly correlated with $\Gamma_s$. We compare our observations  in Figure~\ref{fig:gammaplots} with X-ray-weak NLS1s \citep{WMP04} and PG quasars \citep{Porquet04}; the 0.3--2 keV {\em XMM-Newton} observations of \citeauthor{Porquet04} are converted to 0.5--2 keV using their derived spectral slopes. Eddington ratios are derived using $L_{\rm H\alpha}$ to estimate $L_{5100\AA}$ \citep{GH05,GH07b}, and assuming $L_{\rm bol} = 9 \lambda L_{\rm 5100 \AA}$
, which has a typical scatter of $\approx 0.4$ dex \citep{Ho08}. We use optical luminosities to avoid potentially spurious correlations between $\Gamma_s$ and $L_{\rm X}$ which may arise because we are more sensitive to soft sources with {\em Chandra}. 

These results are consistent with other published results. For instance, \cite{Shemmer06} (and references therein) argue for the hard X-ray spectral index $\Gamma_{\rm h}$ (defined for energies greater than 2 keV) depending primarily on the accretion rate. They find that $\Gamma_{\rm h}$ increases with increasing $L_{\rm bol}/L_{\rm Edd}$ (ranging from 0.05 to 1.0), qualitatively similar to our results. This has also been discussed by, for instance, \cite{BB98} and \cite{LY99}. A typical explanation for this correlation invokes a high accretion rate driving up the disk temperature, producing more soft disk photons which could Compton cool the corona, reducing the hard X-ray emission and thus steepening the X-ray spectral index \citep{HM93,PDO95}.

NLS1s also exhibit pronounced X-ray variability on short timescales \citep{BBF96,Leighly99a}. Our data do not have long enough exposures to make any meaningful variability measurements, although we can use archival data to investigate long-term variability. In Paper I four objects had {\em ROSAT} All-Sky Survey (RASS) detections, but only GH01 showed significant variability (factor of $\approx 5$) over this $\approx 10$ yr timescale in soft X-rays. Two objects in our sample have archival RASS data; GH12 is currently $\approx 2$ times brighter whereas GH17 is $\approx 4$ times brighter. Thus half of the \citeauthor{GH04} sample detected by {\em ROSAT} exhibit factor of few variability over decadal timescales, with the other half limited to small-amplitude ($<$50\%) variability. 

\subsection{X-ray-to-Optical Flux Ratio}

\begin{figure}
\epsscale{1.0}
\plotone{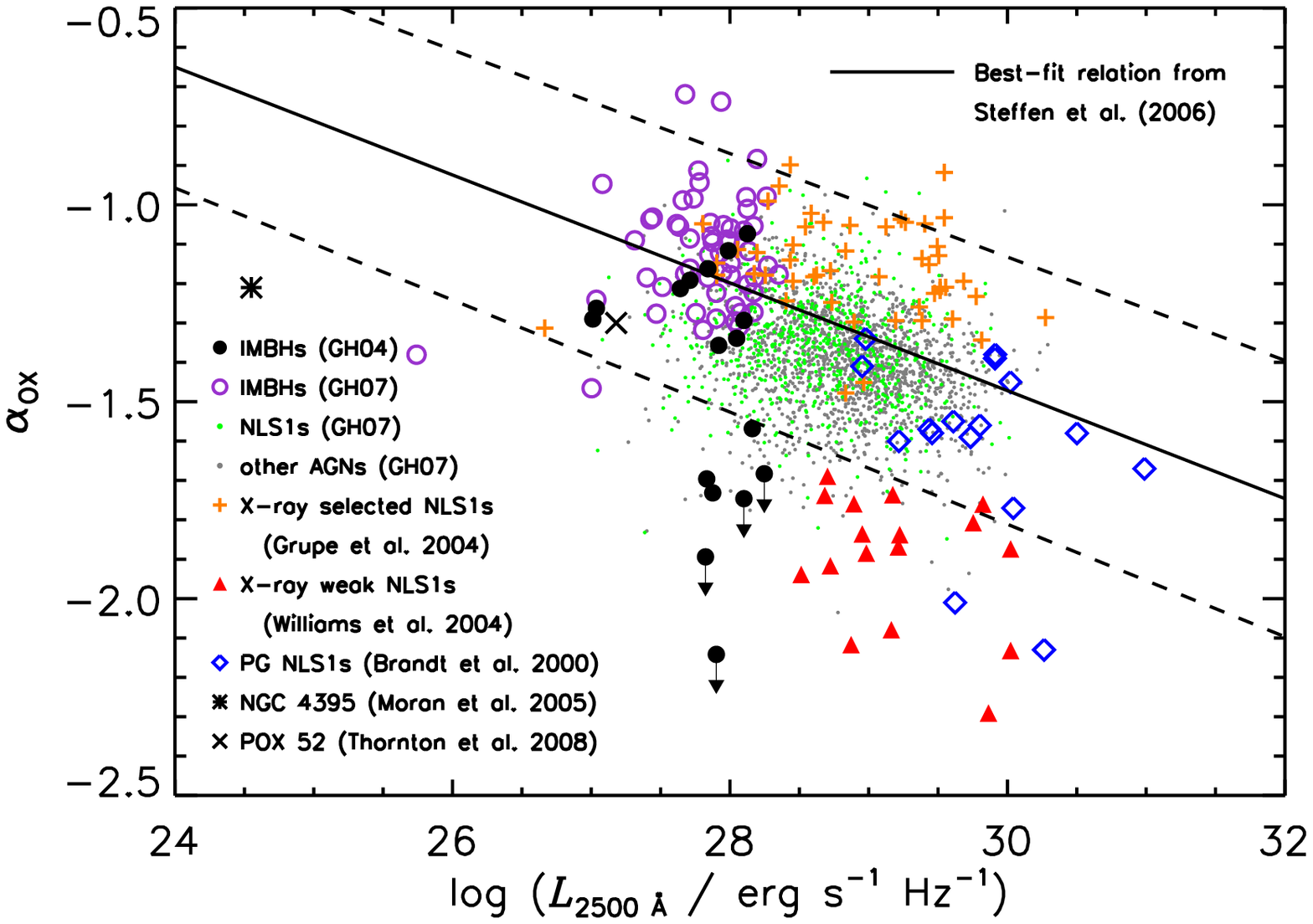}
\caption{X-ray--optical spectral index $\alpha_{\rm ox}$ vs. monochromatic 2500 \AA \ luminosity, in units of erg s$^{-1}$ Hz$^{-1}$. Black circles are the \cite{GH04} objects detected by {\em Chandra}, where $L_{2500\AA}$ is determined from \ha \ measurements \citep{GH05,GH07b}. We also plot the sample of 55 {\em ROSAT}-detected, intermediate-mass BHs \citep[{\it purple open circles}]{GH07c}, the full sample of {\em ROSAT}-detected, low-redshift ($z<0.35$), type 1 AGNs from SDSS \citep[{\it gray points}]{GH07b}, NLS1s from that same sample ({\it green points}), X-ray-weak NLS1s \cite[{\it red triangles}]{WMP04}, PG NLS1s ({\it blue open diamonds}), X-ray-selected NLS1s \cite[{\it orange pluses}]{Grupe04}, NGC\,4395 \citep[{\it asterisk}]{Moran05}, and POX\,52 \citep[{\it cross}]{Thornton08}. The best-fit $L_{2500\AA}$ -- $\alpha_{\rm ox}$ relation from \cite{Steffen06} is also included ({\it solid line}; 1$\sigma$ width of line given by {\it dashed lines}).
}
\label{fig:aox}
\end{figure}

The ratio of the optical-to-X-ray flux is an important broadband diagnostic of the broader spectral energy distribution. To characterize this ratio, we use $\alpha_{\rm ox}$, the slope of a hypothetical power law extending from 2500 \AA \ to 2 keV \citep{Tananbaum79}. We adopt the following definition: $\alpha_{\rm ox} \equiv -0.3838 \log (f_{2500\AA}/f_{\rm 2 \ keV}) $ \citep{Strateva05}. To obtain a flux density at 2500 \AA, we use \ha \ measurements from \cite{GH07b} to determine the AGN flux density at 5100 \AA \ \citep{GH05}. We then assume a power-law optical continuum such that $f_\lambda \propto \lambda^{-\beta}$, with an average $\langle \beta \rangle = 1.56 
\pm 0.1 $ \citep{VandenBerk01,GH07c}, to calculate $f_{2500\AA}$. This differs from Paper I in that we do not use the measured $L_{5100\AA}$, which is potentially affected by galaxy starlight. In Figure~\ref{fig:aox} we plot $\alpha_{\rm ox}$ vs. the monochromatic luminosity at 2500 \AA, a well-known correlation \citep{AT82,Bechtold03}. We include the X-ray-weak NLS1 sample of \cite{WMP04}, X-ray-selected NLS1s \citep{Grupe04}, and the PG quasars that are classified as NLS1s according to FWHM$_{\rm H\beta}$ \citep{BG92}, with $\alpha_{\rm ox}$ given by \cite{BLW00}. We also include upper limits for our non-detections, with the exception of GH19, which is not considered an intermediate-mass BH candidate according to the revised detection algorithm of \cite{GH07c}. Our sample continues to agree reasonably well with extrapolations to lower luminosity (and mass). We note that the \cite{WMP04} sample is likely X-ray-weak as a result of intrinsic absorption \citep{BLW00}.

In a larger sample of 174 SDSS-selected intermediate-mass BHs in active galaxies, 55 are detected by {\em ROSAT} \citep{GH07c}. If we do not restrict ourselves to intermediate-mass BHs but consider all SDSS-selected AGNs \citep{GH07b} with cross-identifications in both the SDSS and the RASS, then our sample grows to 2235 objects (of which 658 are NLS1s with FWHM$_{\rm H \beta} < 2000$ km s$^{-1}$). We include all these samples in Figure~\ref{fig:aox}. For the SDSS-RASS sample, we estimate $\alpha_{\rm ox}$ by converting the {\em ROSAT} 0.1--2.4 keV counts to 0.5--2 keV using WebPIMMS, assuming an absorbed power law with index $\Gamma_s = 2$, Galactic extinction of $\log(N_{\rm H}) = 20.27$ (the median of the \citeauthor{GH07c} sample), and redshift $z=0.19$ (median of the whole sample). The monochromatic luminosity $L_{\rm 2500 \AA}$ is once again estimated from the \ha \ emission line.

\begin{figure}
\epsscale{1.0}
\plotone{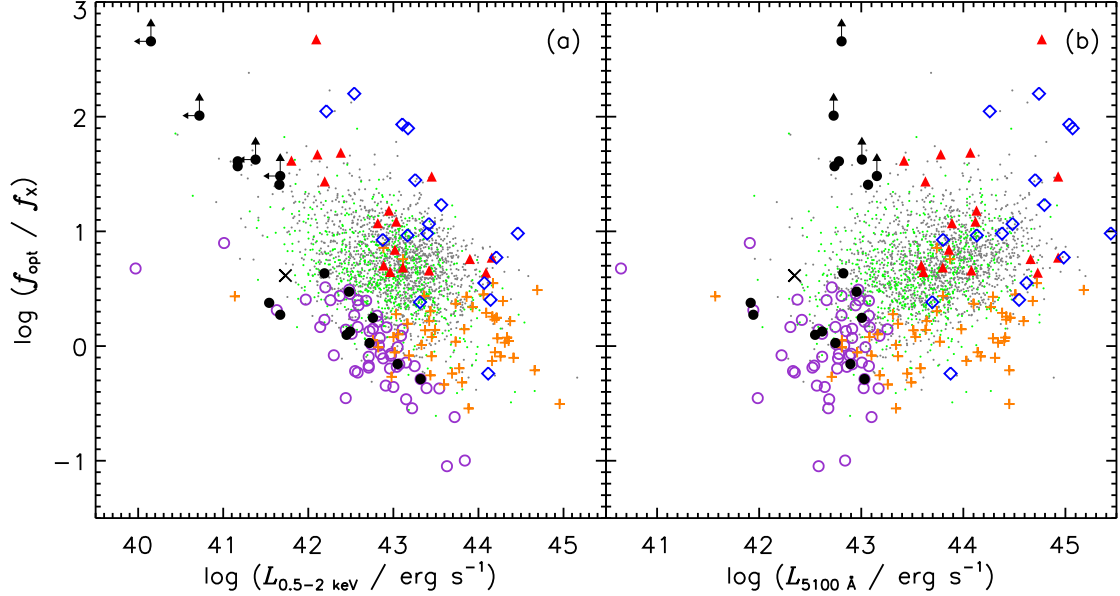}
\caption{Optical-to-X-ray flux ratio vs. (a)  X-ray luminosity, and (b) optical luminosity. X-ray flux is defined over 0.5--2 keV. Optical flux is defined as $\lambda f_\lambda$ at $\lambda=5100$ \AA, determined from \ha \ emission-line measurements \citep{GH05}. Symbols and conventions as in Figure~\ref{fig:aox}.
NGC\,4395 is not shown as it has very low X-ray and optical luminosity.}
\label{fig:foptfx}
\end{figure}

Although intermediate-mass BHs follow extrapolations of $\alpha_{\rm ox}$ to lower mass, in general this sample appears to be X-ray-bright. This can be seen by plotting the direct optical-to-X-ray flux ratio (where $f_{\rm opt} = \lambda f_\lambda$ at 5100 \AA, and $f_ {\rm X}$ is the 0.5--2 keV flux) as a function of both the optical luminosity and the X-ray luminosity (Figure~\ref{fig:foptfx}). The intermediate-mass BH sample is at the lower end of the optical luminosity range exhibited by AGNs in SDSS, but has a comparable X-ray luminosity as other BHs many orders of magnitude higher in mass. This is an important result since these two samples are drawn from the same SDSS-RASS catalog and observed with the same instrument.
Furthermore, although the bolometric corrections and BH mass estimators used suffer from large scatter and systematic uncertainty (for instance, with the geometry of the broad-line region), all RASS objects were treated uniformly.
The systematically higher $\alpha_{\rm ox}$ for the intermediate-mass BHs is thus a real effect (supported by our {\em Chandra} observations). 
The optical-to-X-ray flux ratio is similar to that of \cite{Grupe04}, who {\em selected} their objects based on X-ray flux (and therefore represent the high-X-ray end of the distribution of NLS1s). Our sample has no such selection, is drawn from a uniform SDSS parent sample based solely on $M_{\rm BH}$, and yet displays significantly lower optical-to-X-ray flux ratios than NLS1s drawn from the same parent sample. 

\begin{figure}
\epsscale{1.0}
\plotone{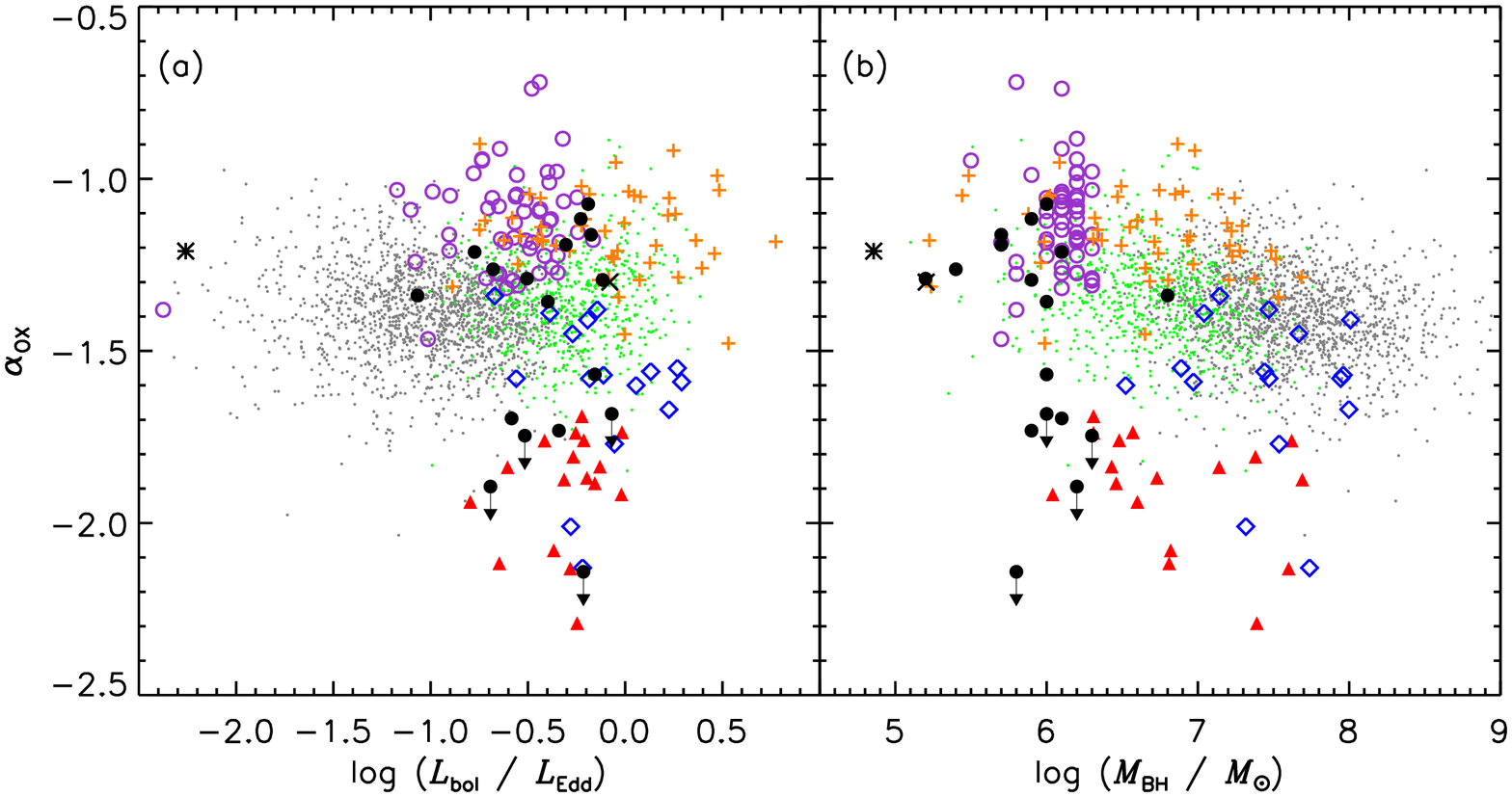}
\caption{X-ray--optical spectral index $\alpha_{\rm ox}$ vs. (a) Eddington ratio; and (b) $M_{\rm BH}$. The bolometric luminosity is assumed to follow $L_{\rm bol} = 9 L_{\rm 5100 \AA}$ \citep{Ho08}, with $L_{\rm 5100 \AA}$ estimated from \ha \ emission-line measurements \citep{GH05}. The X-ray flux is defined over 0.5--2 keV. The BH masses in our sample are estimated via virial techniques \citep{GH05}. Symbols and conventions as in Figure~\ref{fig:aox}.
Masses for PG NLS1s, NGC\,4395 and POX\,52 are given by \cite{VP06}, \cite{Peterson05}, and \cite{Barth04}, respectively.}
\label{fig:aox_alt}
\end{figure}

What could be driving the relative X-ray loudness of the low-mass sample?
The three obvious physical parameters are $L_{\rm bol}$, $L_{\rm bol}/L_{\rm Edd}$, and $M_{\rm BH}$. As is the case for the spectral index, it has been suggested that $\alpha_{\rm ox}$ depends on the accretion rate of the BH \citep{Kelly08}. Given the observed correlation between $\Gamma_s$ and $L_{\rm bol}/L_{\rm Edd}$ (Figure~\ref{fig:gammaplots}), it is reasonable to expect a similar correlation with respect to $\alpha_{\rm ox}$. In Figure~\ref{fig:aox_alt} we plot $\alpha_{\rm ox}$ against $L_{\rm bol}/L_{\rm Edd}$ and $M_{\rm BH}$. We clearly see, however, that accretion rate, as defined by the optical luminosity, is {\em not} a driver of $\alpha_{\rm ox}$ for the bulk of the AGNs, at least over this range in Eddington ratios, which span roughly a factor of 100 from $L_{\rm bol}/L_{\rm Edd} \approx 10^{-2}$ to $1$. If we assume that the intrinsic scatter is more dominant than the error on individual measurements, that the scatter is symmetric, and we force $\chi^2$ per degree of freedom to equal 1, then using the fit routine {\tt fitexy} results in $\alpha_{\rm ox} \propto (0.002 \pm 0.007) \log (L_{\rm bol}/L_{\rm Edd})$ for the large SDSS AGN sample, consistent with no dependence. We also find only a weak correlation $\alpha_{\rm ox} \propto -(0.052 \pm 0.005) \log (M_{\rm BH})$. Because we estimate $L_{\rm 2500 \AA}$ and ultimately $L_{\rm bol}$ based on the \ha \ emission line, Figure~\ref{fig:aox_alt} also suggests that $\alpha_{\rm ox}$ is independent of $L_{\rm UV}/L_{\rm Edd}$. The higher $\alpha_{\rm ox}$ for the intermediate-mass BHs is therefore more of a discontinuous jump from the main AGN population, rather than a smooth correlation.

These results are somewhat at odds with \cite{Kelly08}, who find $L_{\rm UV}/L_{\rm Edd}$ and $M_{\rm BH}$ to be correlated with $\alpha_{\rm ox}$ over a wide range in mass and accretion rate. It should be noted that their range of $L_{\rm bol}/L_{\rm Edd}$ is larger than ours and thus perhaps they see a correlation that we do not probe. Their sample also spans a much wider range in redshift. Furthermore, we employ different correlations to convert from emission-line fluxes to continuum fluxes. In our sample, however, we treat all the SDSS-RASS objects in the same manner; therefore the higher $\alpha_{\rm ox}$ for intermediate-mass BHs is not an artifact. 

Our results are consistent with Paper I, in which $\alpha_{\rm ox}$ was independent of $L_{\rm bol}/L_{\rm Edd}$ for the {\em Chandra} sample; this now appears to be a more general property of AGNs, with two distinct subgroups. The bolometric fraction of the X-ray emission increases with decreasing UV/optical continuum strength, independent of the Eddington ratio. This property can be explained via disk-corona models, where soft disk photons cool the X-ray-emitting corona via Compton cooling, as well as thermally reprocessing some fraction of hard X-rays \citep{HM93}. Of course, the lack of dependence on $L_{\rm bol}/L_{\rm Edd}$ extends only so far; at low enough Eddington ratio, the accretion flow is thought to transition from an optically thick, geometrically thin disk \citep{SS73} to a radiatively inefficient, optically thin, and geometrically thick disk \citep{Ho99,Ho08,Quataert01,Narayan05}. At this transition, there will surely be a sharp change in $\alpha_{\rm ox}$. \cite{Ho08} sees evidence for such a transition at around $\log (L_{\rm bol}/L_{\rm Edd}) \approx -3$. In the case of low-ionization nuclear emission-line regions (LINERs), thought to be a result of radiatively inefficient accretion \citep{Ho08}, $\alpha_{\rm ox}$ falls below a naive extrapolation of the \citeauthor{Steffen06} best fit to lower luminosity \citep{Maoz07}. Perhaps an analogous transition is occurring for intermediate-mass BHs. 

\subsection{Possible Evidence for Slim Disks}

Given that our low-BH-mass sample exhibits a systematically flatter $\alpha_{\rm ox}$ at a given optical luminosity and Eddington ratio, it may indicate an important physical distinction in the accretion flow associated with the low BH mass. One possibility is that intermediate-mass BHs form {\em slim disks} \citep{Abramowicz88}, an accretion disk model that has historically been invoked for nearly-Eddington or super-Eddington accretion flows (the possibility of super-Eddington flows is discussed by \citealt{OM07} and \citealt{Ohsuga05}). In such disks, the very high temperature and luminosity of the inner accretion disk causes the geometrically thin disk to become inflated at small radii, creating a pronounced atmospheric structure. The transition radius between the inner slim disk and the outer thin disk increases with increasing $L_{\rm bol}/L_{\rm Edd}$ \citep{Bonning07}. At small radii, the accretion becomes radiatively inefficient because of photon-trapping effects and radial advection of material sets in. Locally radiated flux within this transition radius (and the associated effective temperature) may then be depressed when observing this disk. 
Slim-disk models have luminosities and effective temperatures (at fixed BH mass) that are nearly independent of accretion rate,
since any extra energy emitted as a result of higher $L_{\rm bol}/L_{\rm Edd}$ falls directly into the BH \citep{Wang99,Mineshige00}. The expected $\alpha_{\rm ox}$ is close to $-1$, flatter than typical AGNs \citep{Mineshige00}. The relatively stronger X-ray emission may be a result of this inflated inner-disk atmosphere, with plenty of X-ray-emitting, hot and diffuse gas, coupled with depressed optical emission due to the radial advection of energy at small radii. 
 
Slim disks are thought to become important above $L_{\rm bol}/L_{\rm Edd} \approx 0.3$ \citep{Bonning07}; we indeed see a flare up in $\alpha_{\rm ox}$ above such an accretion rate in Figure~\ref{fig:aox_alt}. The low-redshift, {\em ROSAT}-detected AGNs, including classical NLS1s, exhibit a remarkably constant $\alpha_{\rm ox}$, whereas {\em ROSAT}-detected, intermediate-mass BHs have systematically higher values. The \citeauthor{Grupe04} NLS1 sample also lies above the main AGN population. Given that the \citeauthor{Grupe04} objects exhibit the highest accretion rates of objects considered here (a fraction of which are super-Eddington), it seems plausible that these are genuine slim disks. The similarity in $\alpha_{\rm ox}$ between intermediate-mass BHs and the \citeauthor{Grupe04} NLS1s suggests that slim disks might be important at intermediate mass as well. As \cite{Mineshige00} discuss, the prediction that $\alpha_{\rm ox} \approx -1$ for slim disks is a potential problem, since most NLS1s have $\alpha_{\rm ox} \approx -1.5$. As is clear from Figure~\ref{fig:aox_alt}, however, we measure values close to $-1$ 
for intermediate-mass BHs and super-Eddington NLS1s; the mean $\alpha_{\rm ox}$ is $-1.12 \pm 0.02$ for the {\em ROSAT} intermediate-mass BHs, and $-1.18 \pm 0.01$ for the \citeauthor{Grupe04} sample, whereas the mean $\alpha_{\rm ox}$ is $-1.39 \pm 0.01$ for the low-redshift SDSS-RASS AGNs, and $-1.36 \pm 0.01$ for the low-redshift SDSS-RASS NLS1s. 

\begin{figure}
\epsscale{1.0}
\plotone{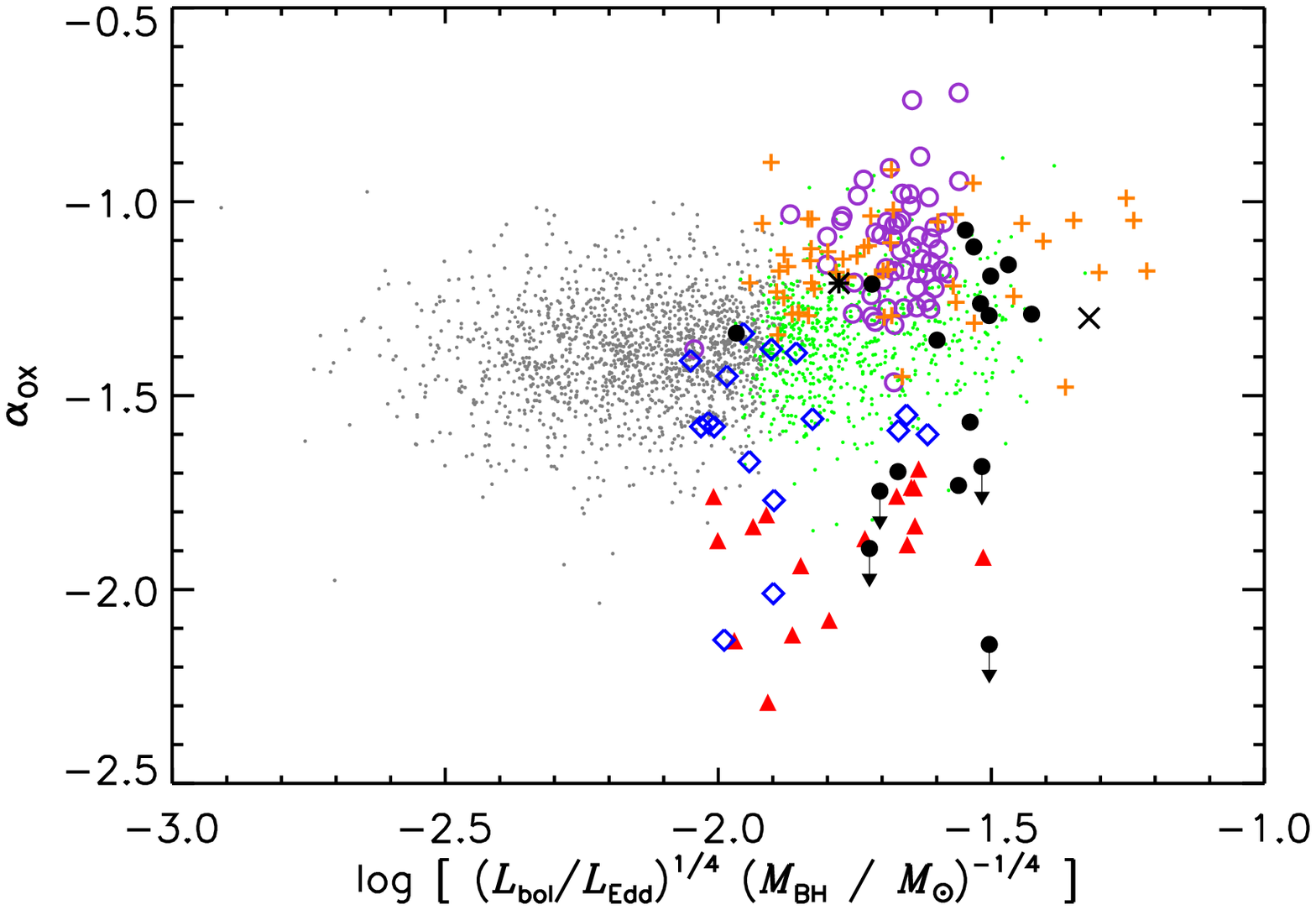}
\caption{X-ray--optical spectral index $\alpha_{\rm ox}$ vs. effective temperature of the disk \citep{FKR92}. The bolometric luminosity is assumed to follow $L_{\rm bol} = 9 L_{\rm 5100 \AA}$ \citep{Ho08}, with $L_{\rm 5100 \AA}$ estimated from \ha \ emission-line measurements \citep{GH05}. The BH masses are estimated via virial techniques \citep{GH05}. Symbols and conventions as in Figure~\ref{fig:aox}.
Masses for PG NLS1s, NGC\,4395 and POX\,52 are given by \cite{VP06}, \cite{Peterson05}, and \cite{Barth04}, respectively.}
\label{fig:aox2}
\end{figure}

As intriguing as these results might be, we must remember the inherent uncertainty associated with these BH-mass and Eddington-ratio estimates. This makes comparisons to other accreting BH systems, such as stellar BH binaries with more accurately measured mass functions and observed state changes, difficult at best. The important result, however, remains: intermediate-mass BHs exhibit a distinct spectral energy distribution compared to higher-mass NLS1s. Slim disks provide an interesting framework for the interpretation of this result. The change in energy distribution is clearly driven by some combination of BH mass and Eddington ratio, which unfortunately we cannot fully disentangle with our current data. For example, in Figure~\ref{fig:aox2} we plot $\alpha_{\rm ox}$ versus $(L_{\rm bol}/L_{\rm Edd})^{1/4} M_{\rm BH}^{-1/4}$, which is proportional to the disk effective temperature \citep{FKR92}. The higher $\alpha_{\rm ox}$ objects are those with high disk temperatures, but not all high-temperature objects exhibit a high $\alpha_{\rm ox}$ (i.e. most NLS1s). Further study is clearly warranted, ideally with a comprehensive, homogenous, and detailed survey. Complete spectral energy distributions are needed to quantify the varying X-ray bolometric corrections between the different AGN classes.

\section{Summary}
\label{sec:sum}

We present X-ray observations of the remaining intermediate-mass BHs found by \cite{GH04} and not observed by \cite{GH07a}. We detect 5 out of 8 objects in 5 ks observations with {\em Chandra}. The mean observed properties, such as hardness ratio and soft photon index $\Gamma_s$, are similar to the initial sample; we continue to find a range of indices $1 < \Gamma_s < 3$, consistent with previous AGN studies. Only 2 objects have sufficient counts to extract reliable spectra, and both are well fit with simple absorbed power-law models. The resulting $\chi^2$ values do not justify any additional components, such as a soft excess, although this may be a function of relative inefficiency of detecting counts above 2 keV in short exposures. 
The soft photon index continues to be correlated strongly with X-ray luminosity and Eddington ratio, while the BH mass remains a poor indicator of the X-ray spectral slope. 

Although the \citeauthor{GH04} sample shares many characteristics with classical NLS1s, there are important differences between the two. In particular, the X-ray-to-optical flux index $\alpha_{\rm ox}$ of these intermediate-mass AGNs is flatter, similar to NLS1s radiating near or above the Eddington limit. This may be evidence for a change in the accretion structure of such systems, perhaps due to the formation of a slim disk instead of a classical thin disk. There appears to be a sharp transition; within the two groups (intermediate-mass AGNs and super-Eddington NLS1s vs. normal AGNs and classical NLS1s) $\alpha_{\rm ox}$ is independent of the Eddington ratio. Additionally, we do not see evidence for very steep soft photon indices, as suggested by \cite{BBF96} and \cite{Porquet04}, despite the very low BH masses. 

As was shown in the pilot study, the feasibility of detecting intermediate-mass BHs with short {\em Chandra} exposures is clearly established. Such observations are a vital component to understanding the broad spectral energy distribution and behavior of BHs in this previously unobserved intermediate-mass regime, and to properly measure and calibrate their bolometric luminosities. Our single-epoch, short exposures do not allow us to study X-ray variability on short timescales, known to be very pronounced in NLS1s, although we note that variability of factors of a few can be seen when compared with $\approx 10$ yr old {\em ROSAT} archival data. Further X-ray observations of such objects would help to clarify these issues. 

\acknowledgements

We thank an anonymous referee for thoughtful comments. L.-B.D. would like to thank Eliot Quataert and J.E.G. Peek for helpful and insightful discussions. Support for this work was provided by NASA through grant SAO 08700135 issued by the {\em Chandra} X-ray Observatory Center, which is operated by the Smithsonian Astrophysical Observatory on behalf of NASA under contract NAS8-03060, and through grant HST-GO-11130.01-A from the Space Telescope Science Institute, which is operated by the Association of Universities for Research in Astronomy, Inc., for NASA, under contract NAS5-26555. 


\clearpage
\thispagestyle{empty}

\begin{landscape}
\begin{deluxetable}{lcccccccccccccc}
\centering
\tablewidth{0pt}
\tabletypesize{\scriptsize}
\tablecaption{X-Ray Properties}
\tablehead{\colhead{ID} & \colhead{$D_L$} & \colhead{$\log N_{\rm H}$} & \colhead{$C_s$} & \colhead{$C_h$} & \colhead{$H$} & \colhead{$\Gamma_{\rm HR}$} & \colhead{$\log f_s$} & \colhead{$\log f_h$} & \colhead{$\log L_s$} & \colhead{$\log L_h$} & \colhead{$\log L_{\rm H\alpha}$} & \colhead{$\alpha_{\rm ox}$} & \colhead{$\log M_{\rm BH}$} & \colhead{Ref.} \\ 
\colhead{(1)} & \colhead{(2)} & \colhead{(3)} & \colhead{(4)} & \colhead{(5)} & \colhead{(6)} & \colhead{(7)} & \colhead{(8)} & \colhead{(9)} & \colhead{(10)} & \colhead{(11)} & \colhead{(12)} & \colhead{(13)} & \colhead{(14)} & \colhead{(15)}}
\startdata
GH01 \dotfill & 343 & 20.59 & 0.183 $\pm$ 0.007 & 0.0258 $\pm$ 0.003 & $-0.75 \pm 0.02$  & 2.5 $\pm$ 0.1 & $-12.10^{+0.05}_{-0.07}$ &  $-12.35^{+0.05}_{-0.08}$ &  $43.05^{+0.05}_{-0.07}$ &  $42.81^{+0.05}_{-0.07}$ & 41.44 & $-1.12$ & 5.9 & 1\\
GH02 \dotfill & 	127	& 20.60 & 0.0437 $\pm$ 0.003	& 0.0104 $\pm$ 0.002 & $-0.61 \pm 0.04$ & 2.2 $\pm$ 0.2 & $-12.75^{+0.12}_{-0.20}$ & $-12.82^{+0.13}_{-0.24}$ & $41.54^{+0.12}_{-0.20}$ & $41.47^{+0.12}_{-0.20}$ & 40.31 & $-1.29$ & 5.2 & 1 \\
GH03 \dotfill & 466 & 20.88 & 0.0013 $\pm$ 0.0008 & 0.0004 $\pm$ 0.0006 & $-0.5 \pm 1$ & $\approx 1.9$ & $\approx -14.25$ & $\approx -14.19$ & $\approx 41.17$ & $\approx 41.23$ & 41.31 & $-1.73$ & 5.9 & 2 \\
GH04 \dotfill &	189	& 20.61 & 0.176 $\pm$ 0.006 & 0.0415 $\pm$ 0.003 & $-0.62 \pm 0.01$ & 2.2 $\pm$ 0.1 & $-12.14^{+0.06}_{-0.08}$ &  $-12.22^{+0.07}_{-0.05}$ &  $42.49^{+0.06}_{-0.08}$ & $42.42^{+0.06}_{-0.08}$& 41.12 & $-1.19$ & 5.7 & 1 \\
GH05 \dotfill &	331	& 20.35 & $<$ 0.00111 & $<$ 0.0001 & \ldots & \ldots & $< -14.97$  & $< -14.37$  & $<$ 40.15 & $<$ 40.75 & 41.34 & $< -2.14$ & 5.8 & 1\\
GH06 \dotfill & 455 & 20.51 & 0.016 $\pm$ 0.002 & 0.004 $\pm$ 0.001 & $-0.58 \pm 0.06$ & 1.9 $\pm$ 0.2 & $-13.21^{+0.05}_{-0.06}$ & $-13.19^{+0.10}_{-0.13}$ & 42.19$^{+0.05}_{-0.06}$ & 42.21$^{+0.10}_{-0.13}$ & 41.36 & $-1.36$ & 6.0 & 2 \\
GH07 \dotfill &	427	& 20.59 & 0.0245 $\pm$ 0.003	& 0.00362 $\pm$ 0.001 & $-0.74 \pm 0.03$ & 2.5 $\pm$ 0.1 & $-12.89^{+0.22}_{-0.65}$ &  $-13.13^{+0.16}_{-0.84}$ &  $42.45^{+0.22}_{-0.65}$ & $42.21^{+0.22}_{-0.65}$& 41.04 & $-1.21$ & 6.1 & 1 \\
GH08 \dotfill &	364	& 20.37 & 0.0811 $\pm$ 0.004	& 0.00942 $\pm$ 0.002 & $-0.79 \pm 0.03$ & 2.7 $\pm$ 0.2 & $-12.44^{+0.09}_{-0.12}$ &  $-12.79^{+0.08}_{-0.12}$ &  $42.76^{+0.09}_{-0.12}$ & $42.41^{+0.09}_{-0.12}$& 41.57 & $-1.29$ & 5.9 & 1 \\
GH09 \dotfill & 945 & 20.35 & $<$ 0.0013 & $<$ 0.0015 & \ldots & \ldots & $< -14.34$ & $< -13.61$ & $<$ 41.67 & $<$ 42.42 & 41.74 & $< -1.68$ & 6.0 & 2 \\
GH10 \dotfill &	363	& 20.27 & 0.00727 $\pm$ 0.001 & 0.00747 $\pm$ 0.001 & 0.014 $\pm$ 0.05 & 1.0 $\pm$ 0.1 & $-13.53^{+0.42}_{-2.46}$ &  $-12.92^{+0.49}_{-3.08}$ &  $41.66^{+0.42}_{-2.46}$ & $42.28^{+0.42}_{-2.46}$& 41.64 & $-1.57$ & 6.0 & 1\\
GH11 \dotfill &	365	& 20.24 & 0.00317 $\pm$ 0.001 & 0.00145 $\pm$ 0.001 & $-0.37 \pm 0.10$ & 1.7 $\pm$ 0.1 & $-14.04^{+0.82}_{-1.94}$ &  $-13.81^{+0.83}_{-2.17}$ &  $41.17^{+0.82}_{-1.94}$ & $41.40^{+0.82}_{-1.94}$& 41.26 & $-1.70$ & 6.1 & 1 \\
GH12 \dotfill & 487 & 20.31 & 0.193 $\pm$ 0.007 & 0.023 $\pm$ 0.003 & $-0.78 \pm 0.01$ & 2.6 $\pm$ 0.1 & $-12.13^{+0.02}_{-0.02}$ & $-12.50^{+0.05}_{-0.06}$ & 43.32$^{+0.02}_{-0.02}$ & 42.95$^{+0.05}_{-0.06}$ & 41.60 & $-1.07$ & 6.0  & 2\\
GH14 \dotfill &	121	& 20.40 & 0.0568 $\pm$ 0.004	& 0.0114 $\pm$ 0.002 & $-0.67 \pm 0.03$ & 2.3 $\pm$ 0.1 & $-12.58^{+0.11}_{-0.24}$ &  $-12.69^{+0.15}_{-0.18}$ &  $41.67^{+0.11}_{-0.24}$ & $41.55^{+0.11}_{-0.24}$& 40.34  & $-1.26$ & 5.4 & 1 \\
GH15 \dotfill & 595 & 20.16 & $<$ 0.0016 & $<$ 0.0020 & \ldots & \ldots & $< -14.25$ & $< -13.48$ & $<$ 41.38 & $<$ 42.15 & 41.57 & $< -1.75$ & 6.3 & 2 \\
GH16 \dotfill & 308 & 20.37 & $<$ 0.0013 & $<$ 0.0016 & \ldots & \ldots & $< -14.34$ & $< -13.58$ & $<$ 40.72 & $<$ 41.48 & 41.25 & $< -1.89$ & 6.2 & 2\\
GH17 \dotfill & 453 & 20.51 & 0.054 $\pm$ 0.004 & 0.010 $\pm$ 0.002 & $-0.70 \pm 0.03$ & 2.3 $\pm$ 0.1 & $-12.67^{+0.03}_{-0.03}$ & $-12.83^{+0.08}_{-0.10}$ & 42.72$^{+0.03}_{-0.03}$ & 42.56$^{+0.08}_{-0.10}$ & 41.27 & $-1.16$ & 5.7 & 2\\
GH18 \dotfill & 886 & 20.61 & 0.008 $\pm$ 0.002 & 0.0013 $\pm$ 0.0008 & $-0.73 \pm 0.13$ & 2.4 $\pm$ 0.6 & $-13.49^{+0.10}_{-0.13}$ & $-13.73^{+0.21}_{-0.43}$ & 42.48$^{+0.10}_{-0.13}$ & 42.25$^{+0.21}_{-0.43}$ & 41.51 & $-1.34$ & 6.8 & 2 \\
GH19 \dotfill &	154	& 20.58 & $<$ 0.00117 & $<$ 0.00113 & \ldots & \ldots & $< -14.94$ & $< -14.35$ & $<$ 39.52 & $<$ 40.10 & & & & 1
\enddata
\tablecomments{Col. (1): Identification number from \citealt{GH04}. Col. (2): Luminosity distance, in Mpc, calculated using the observed SDSS redshift and our adopted cosmology. Col. (3): Logarithm of neutral column density, in cm$^{-2}$, from \citealt{DL90} using WebPIMMS. Col. (4): 0.5--2 keV count rate, in counts s$^{-1}$. Col. (5): 2--8 keV count rate, in counts s$^{-1}$. Col. (6): Hardness ratio, where $H \equiv (C_h-C_s)/(C_h+C_s)$. Col. (7): Photon index, where $N(E) \propto E^{-\Gamma_{\rm HR}}$, determined from the hardness ratio (see text). Col. (8): 0.5--2 keV flux, in erg s$^{-1}$ cm$^{-2}$. Col. (9): 2--8 keV flux, in erg s$^{-1}$ cm$^{-2}$. Col. (10): 0.5--2 keV luminosity, in erg s$^{-1}$. Col. (11): 2--8 keV luminosity, in erg s$^{-1}$. Col. (12): \ha \ luminosity \citep{GH07b}, in erg s$^{-1}$. GH19 is not included in the new sample \citep{GH07c}, and thus has no updated \ha \ measurement. Col. (13): Ratio of optical to X-ray flux, where $\alpha_{\rm ox} = -0.3838 \log (f_{\rm 2500\AA}/f_{\rm 2 \ keV})$. See text for details. Col. (14): BH mass \citep{GH07b}. GH19 is not included in the new sample \citep{GH07c}, and thus has no updated BH mass measurement. Col. (15): Reference for original observation: 1) \citealt{GH07a}; 2) this work.
}
\label{tab:prop}
\end{deluxetable}
\clearpage
\end{landscape}

\clearpage
\setlength{\voffset}{0mm}
\begin{deluxetable}{lcccc}
\tablewidth{0pt}
\tabletypesize{\scriptsize}
\tablecaption{Spectral Fits}
\tablehead{\colhead{ID} & \colhead{$\Gamma_s$} & \colhead{Norm.} & \colhead{$\chi_\nu^2$} & \colhead{dof} \\
\colhead{(1)} & \colhead{(2)} & \colhead{(3)} & \colhead{(4)} & \colhead{(5)}}
\startdata
GH12 & 2.40 $\pm$ 0.06 & 3.0 $\pm$ 0.2 & 1.10 & 41 \\
GH17 & 2.52 $\pm$ 0.13 & 1.0 $\pm$ 0.1 & 0.94 & 12
\enddata
\tablecomments{Col. (1): Identification number from \cite{GH04}. Col. (2): Power-law index (where $N(E) \propto E^{-\Gamma_s}$). Col. (3): Normalization at 1 keV in 10$^{-4}$ photons s$^{-1}$ keV$^{-1}$. Col (4): Reduced $\chi^2$. Col. (5): Degrees of freedom.}
\label{tab:spectra}
\end{deluxetable}

\clearpage

\end{document}